\documentclass{svproc}
\usepackage{url}
\usepackage{graphicx}
\usepackage{calrsfs}
\RequirePackage{xspace}
\newcommand{\gevcsq}{\ensuremath{{\mathrm{\,Ge\kern -0.1em V^{2}\!/}c^2}}\xspace}

\DeclareMathAlphabet{\pazocal}{OMS}{zplm}{m}{n}
\def\etal{{\it et al.}}
\def\Mbc{M_{\rm bc}}
\def\invfb{\ensuremath{\mbox{\,fb}^{-1}}\xspace}

\def\Y#1S{\ensuremath{\Upsilon{(#1S)}}\xspace}

\begin{document}
\mainmatter              
\title{Radiative and electroweak penguin decays at $e^{+}e^{-}$ $B$-factories}
\titlerunning{Radiative and EW penguin decays at $B$-factories}  
%
\author{S. Sandilya}
\authorrunning{S. Sandilya} 
%
%
\institute{University of Cincinnati, Cincinnati, Ohio 45221, USA\\
  \email{saurabhsandilya@gmail.com}
}
  
\maketitle              

\begin{abstract}
$B$-meson decays involving radiative and electroweak penguin processes are sensitive probes to new physics beyond the standard model. The Belle experiment recently reported measurements of the inclusive radiative decay $B \to X^{}_{s} \gamma$, and exclusive radiative decays $B \to K^{\ast} \gamma$ and  $B \to K^{0}_{S} \eta \gamma$. A lepton-flavor dependent measurements of angular observables for the decays $B \to K^{\ast} \ell \ell$ by Belle hinted at possible deviation from lepton-flavor-universality. Any departure from lepton flavor universality is essentially accompanied by lepton flavor violation. Recently, lepton-flavor-violating decays $B^{0} \to K^{\ast 0} \mu ^{\pm} e^{\mp}$ are searched at Belle, and stringent limits on their branching fractions are set. The BaBar experiment has searched for the decay $B^{+} \to K^{+} \tau^{+} \tau^{-}$, which comprises third generation of the lepton family. The decays $B \to h \nu \bar{\nu}$ are searched at Belle and obtained upper limits for these decays are close to the standard model predictions.  
\end{abstract}
\section{Introduction}
\label{sec:intro}

The $B$-factories, Belle and BaBar experiments, were located at the interaction region of $e^{+}e^{-}$ asymmetric colliders of KEKB and PEP-II, respectively. These $B$-factories had about a decade long very successful operational period, and recorded a combined data sample over 1.5 ${\rm ab^{-1}}$, which corresponds to more than $1.2 \times 10^{9}$ $B$-meson pairs. In these experiments electron and positron beams collide at the $\Y4S$ resonance, which leads to a clean sample of quantum correlated pairs of $B$-mesons and makes analyses with missing final states straightforward. Also, a low background environment at the $B$-factories enables an efficient reconstruction of the neutral particles. 

In the standard model (SM), flavor changing neutral current (FCNC) processes are forbidden at tree level and proceed via penguin loop or box-diagrams at lowest order. In these loops, non-SM heavy particles can also enter. Thus, FCNC processes involving $b \to s$ quark-level transition are among the most sensitive probes for the new physics (NP) beyond the SM. Recently, $B$-factories provided important measurements involving radiative and electroweak penguin $B$ decays. The measurements from Belle and BaBar experiments reviewed here, are based on the total recorded data sample of 711 \invfb and 424 \invfb, respectively.   

\section{Measurements of $B \to X^{}_{s} \gamma$ and $B \to K^{\ast} \gamma$}
\label{btosg}
The FCNC transition $b \to s \gamma$ proceeds dominantly through electromagnetic penguin diagrams and, is sensitive to NP. The branching fraction (BF) of the decay $B \to X^{}_{s} \gamma$~\cite{cc}, $\pazocal{B}(B \to X^{}_{s} \gamma)$~\cite{{1:ex:pdg},{2:ex:hflav}} is consistent with the SM prediction and constraints NP. The uncertainty in the SM prediction of $\pazocal{B}(B \to X^{}_{s} \gamma)$ is about 7\%~\cite{3:th:btosg}, which is close to the current experimental uncertainties. The upcoming Belle II experiment is expected to further improve the uncertainties in measurement to about 3\%~\cite{4:ex:belle2}. On the other hand, the dominant uncertainty in the SM prediction is due to non-pertubative effects, and it is related to the isospin asymmetry ($\Delta^{}_{0-}$) in the decay $B \to X^{}_{s} \gamma$~\cite{5:th:btosg}. If $\Delta^{}_{0-}$ is found to be zero, then it will lead to the reduction in uncertainty in the SM prediction of $\pazocal{B}(B \to X^{}_{s} \gamma)$. Another interesting observable sensitive to NP is the difference of direct $CP$ asymmetries between the $B^{+}$ and $B^{0}$ mesons: $\Delta A^{}_{CP} = A^{}_{CP}(B^{+} \to X^{+}_{s} \gamma) - A^{}_{CP}(B^{0} \to X^{0}_{s} \gamma)$~\cite{6:th:btosg}. Any significant deviation of $\Delta A^{}_{CP}$ from zero will indicate the presence of NP~\cite{{6:th:btosg},{7:th:btosg},{8:th:btosg}}. 

Recently, Belle reported measurements of $\Delta^{}_{0-}$ and $\Delta A^{}_{CP}$ for the decay $B \to X^{}_{s} \gamma$, where $X^{}_{s}$ is reconstructed from 38 exclusive final states~\cite{9:ex:btosg}. Among these reconstructed modes, 11 are flavor non-specific modes, which are only used for $\Delta^{}_{0-}$ measurements. The result for $\Delta^{}_{0-} = (-0.48 \pm 1.49 \pm 0.97 \pm 1.15) \%$ is found to be consistent with zero, where the first uncertainty is statistical, the second is systematic and the third is due to uncertainty in the BF ratio of the $\Upsilon(4S) \to B^{+}B^{-}$ and $\Upsilon(4S) \to B^{0}\bar{B^{0}}$ decays. This measured value of $\Delta^{}_{0-}$ will be important in improving the theoretical uncertainty in $\pazocal{B}(B \to X^{}_{s} \gamma)$. The obtained $\Delta A^{}_{CP}$ value is $(+3.69 \pm 2.65 \pm 0.76)\%$, which is also consistent with zero as well as with the SM prediction, and hence can be used to constrain NP. 

In another analysis of the exclusive decay $B\to K^{\ast}\gamma$, Belle reported measurements of $\Delta_{0+}$ and $A_{CP}$~\cite{10:ex:btosg}. These BF ratios $\Delta_{0+}$ and $A_{CP}$ provide a strong constraint on NP, as form-factor related uncertainties in the theoretical prediction cancel out~\cite{11:th:btosg}. In this analysis, the first evidence of isospin violation is reported with a significance of 3.1 standard deviations ($\sigma$), with a value of $\Delta_{0+} = (+6.2 \pm 1.5 \pm 0.6 \pm 1.2)\%$, where the third uncertainty is again due to the BF ratio of the $\Upsilon(4S) \to B^{+}B^{-}$ and $\Upsilon(4S) \to B^{0}\bar{B^{0}}$ decays. The results for $A_{CP}(B \to K^{*} \gamma) = (-0.4 \pm 1.4 \pm 0.3)\%$ and, $ \Delta A_{CP} = (+2.4 \pm 2.8 \pm 0.5)\%$ are also reported. All these measurements are consistent with the SM.

\section{Measurement of time-dependent $CP$ asymmetries in $B^{0} \to K^{0}_{S} \eta \gamma$}
\label{btoksetag}

According to the SM, the photon polarization in the $b \to s \gamma$ transition is predominantly left-handed. Right-handed currents can, however, enter in the loop through various NP models and enhancing the right-handed photon polarization. A promising avenue to observe such NP scenarios is the measurement of time-dependent $CP$ violation in a decay of the form $B \to P^{}_{1}P^{}_{2}\gamma$, where $P^{}_{1}$ and $P^{}_{2}$ are scalar or pseudoscalar mesons and the $P^{}_{1}P^{}_{2}$ system is a $CP$ eigenstate~\cite{12:th:btoksetag}. A small mixing-induced $CP$ violation parameter ($\pazocal{S}$) is generated via interference between the $\bar{B^{0}} \to P^{}_{1}P^{}_{2}\gamma_{L(R)}$ and $B^{0} \to P^{}_{1}P^{}_{2}\gamma_{L(R)}$ decays. Thus, the value of $\pazocal{S}$ could be enhanced by the NP related right-handed currents.

\begin{figure}[htb]
  \centering
  \includegraphics[width=0.5\textwidth]{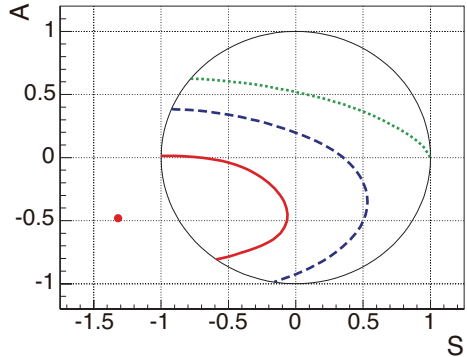}
  \caption{The solid red, dashed blue and dotted green curves show the $1\sigma$, $2\sigma$ and $3\sigma$ confidence contours, respectively. The red dot shows the Belle result, which is is consistent with a null asymmetry within $2\sigma$~\cite{22:ex:btoksetag}. The physical boundary ${\pazocal{S}}^2 + {\pazocal{A}}^2 =1$ is drawn with a thin solid black curve.}
  \label{fig:btoksetag}
\end{figure}

Belle and BaBar, have measured $CP$ violation parameters in the decays $B^{0} \to K^{0}_{s} \pi^{0} \gamma$ (including $K^{0}_{S}\pi^{0}$)~\cite{{13:ex:btoksetag},{14:ex:btoksetag}}, $B^{0} \to K^{0}_{s} \eta \gamma$~\cite{15:ex:btoksetag}, $B^{0} \to K^{0}_{s} \rho^{0} \gamma$~\cite{{16:ex:btoksetag},{17:ex:btoksetag}}, and $B^{0} \to K^{0}_{s} \phi \gamma$~\cite{18:ex:btoksetag}. These results are consistent with SM predictions~\cite{{19:th:btoksetag},{20:th:btoksetag},{21:th:btoksetag}}. First measurement of time-dependent $CP$ asymmetries in the decay $B^{0} \to K^{0}_{s} \eta \gamma$ is reported by Belle~\cite{22:ex:btoksetag}. The obtained parameters are

\begin{eqnarray}
\nonumber
    {\pazocal S} &=& -1.32 \pm 0.77 {\rm (stat.)} \pm 0.36{\rm (syst.)}, \\
\nonumber
    {\pazocal A} &=& -0.48 \pm 0.41 {\rm (stat.)} \pm 0.07{\rm (syst.)},
\end{eqnarray}

that lie outside the physical boundary, defined by ${\pazocal S}^2+{\pazocal A}^2 = 1$, shown in Fig~\ref{fig:btoksetag}. These measurements are consistent with the null-symmetry hypothesis within 2$\sigma$ as well as with SM predictions.       

\section{Angular analysis of $B\to K^{\ast}\ell^{+}\ell^{-}$}
\label{btosll}

The decay $B \to K^{\ast}\ell^{+}\ell^{-}$ involves a quark-level $b \to s\ell^{+}\ell^{-}$ FCNC transition and proceeds via electroweak penguin or a box diagrams. NP particles may enter in these loops, and thus can alter the BF and angular distributions of the final-state particles~\cite{23:th:btosll}. Interestingly, in the recent years, several measurements have shown possible deviations from the SM for a number of decays involving $b \to s\ell^{+}\ell^{-}$ transition~\cite{{24:ex:btosll},{25:ex:btosll},{26:ex:btosll}}. Global fits are performed including experimental and theoretical correlations, and these fits hints at possible lepton flavor universality (LFU) violation~\cite{27:th:btosll}. The angular observables $P^{\prime}_{i}$ for the $B \to K^{\ast}\ell^{+}\ell^{-}$ is introduced in Ref.~\cite{28:th:btosll}, which are considered mostly to be free from form-factor related uncertainities~\cite{29:th:btosll}. And, also if the differences of $P^{\prime}_{i}$ between the muon and the electron modes, $Q^{}_{i} = P^{\prime \mu}_{i} - P^{\prime e}_{i}$ ($i=4,5$), deviates from zero, it would be a signature of NP~\cite{30:th:btosll}.

Belle has reported a measurement of angular observables, $P^{\prime}_{i}$ for both lepton flavors separately and $Q^{}_{i}$, in the decay $B \to K^{\ast}\ell^{+}\ell^{-}$~\cite{31:ex:btosll}. In this measurement, the $B^{+} \to K^{\ast +} \ell \ell$ decays are reconstructed along with $B^{0} \to K^{\ast 0} \ell \ell$ decays, where $K^{\ast +}$ is reconstructed from $K^{+}\pi^{0}$ or $K^{0}_{S}\pi^{+}$ and $K^{\ast 0}$ from $K^{+}\pi^{-}$. The analysis is performed in the four independent bins of $q^{2}$ (invariant mass square of the two leptons). Comprehensively, the results are compatible with SM predictions~\cite{30:th:btosll}. The largest deviation of $2.6\sigma$ from teh SM is observed for $P^{\prime}_{5}$ of the muon modes in the $q^{2} \in (4.0,8.0)\gevcsq$ bin, in the same bin electron modes deviate by $1.3\sigma$, and both combined the deviation is about $2.5\sigma$.

\begin{figure}[htb]
\includegraphics[width=0.5\textwidth]{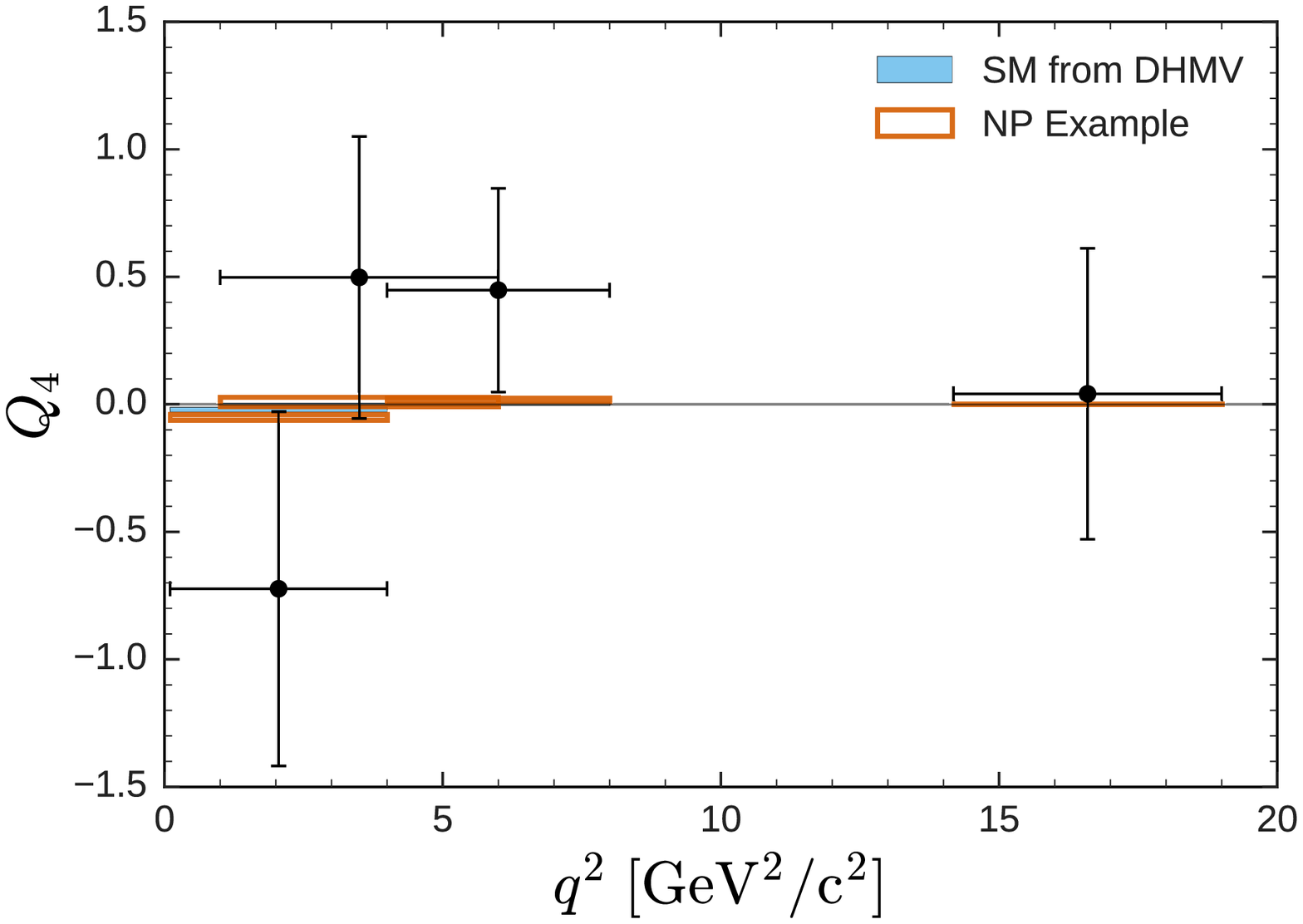}
\includegraphics[width=0.5\textwidth]{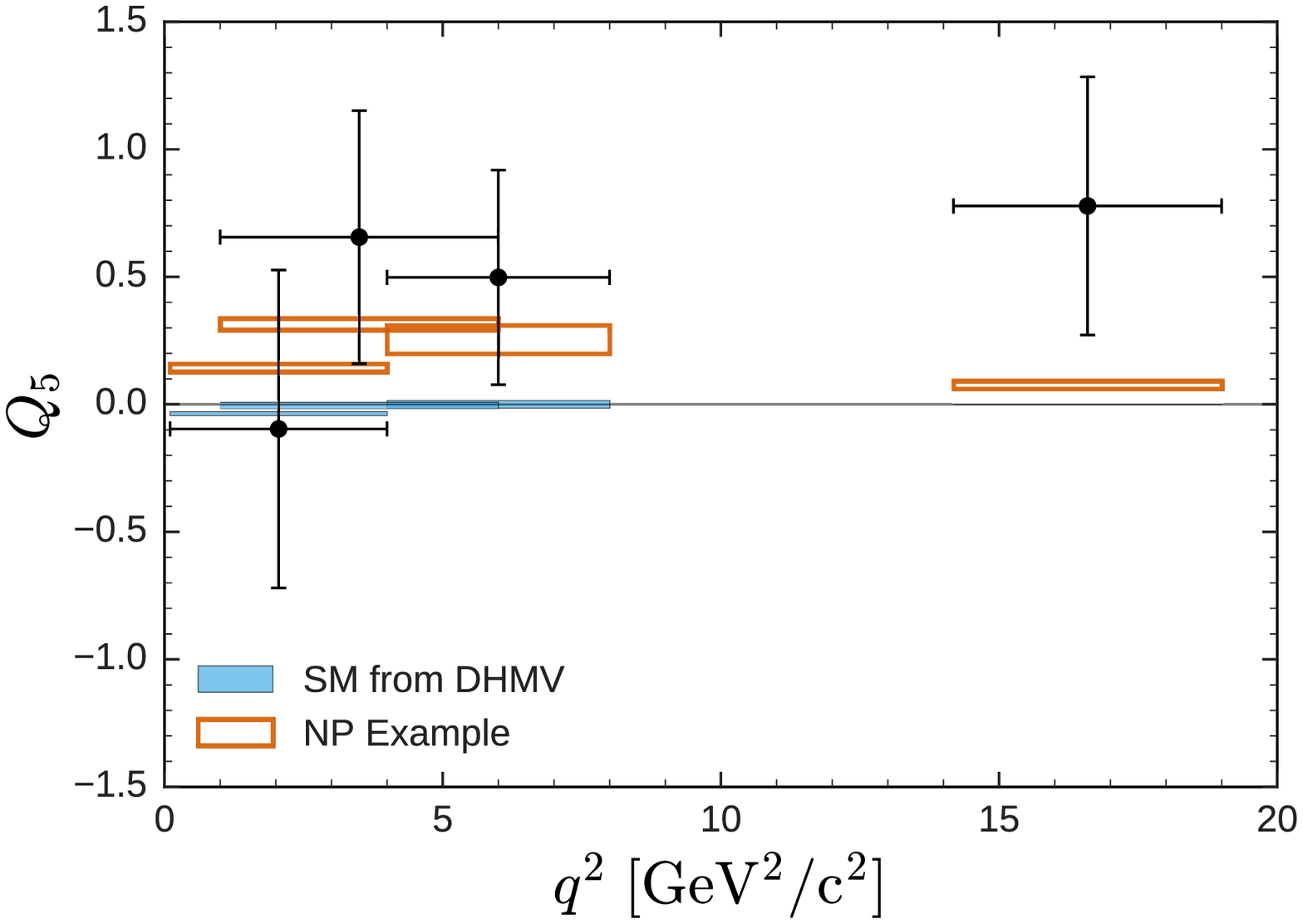}   
\caption{$Q_{4}$ (left) and $Q_{5}$ (right) observables compared with SM and NP scenario~\cite{30:th:btosll}, shown by the cyan filled and brown open boxes~\cite{31:ex:btosll}, respectively.}
\label{fig:btosll:q}
\end{figure}

The $Q^{}_{4,5}$ observable is shown in Fig.~\ref{fig:btosll:q}, where no significant deviation from zero is observed. Global fits including this Belle result hint at LFU violation~\cite{{32:th:btosll},{33:th:btosll}}.

\section{Search for lepton-flavor-violating decays $B^{0}\to K^{\ast 0}\mu^{\pm}e^{\mp}$}
\label{btokstmue}
Over the recent years, measurements of the decays mediated by $b \to s\ell^{+}\ell^{-}$ quark-level transition hint for possible LFU violation, as discussed in Sec.~\ref{btosll}. LFU is an important symmetry of the SM and its violation is usually accompanied by lepton flavor violation (LFV)~\cite{{34:th:btokstmue},{35:th:btokstmue}}. Very recently, Belle has reported a search for the LFV decays $B^{0}\to K^{\ast 0}\mu^{\pm}e^{\mp}$, where $K^{\ast 0}$ is reconstructed from $K^{+}\pi^{-}$~\cite{36:ex:btokstmue}. Backgrounds originating from $e^{+}e^{-}\to q \bar{q}$ ($q=u,d,c,s$ ) continuum processes and other $B$ decays are suppressed with two dedicated neural networks (NN). The $B^{0}\to K^{\ast 0} J/\psi$ decay is used as a control sample. Further, a set of vetoes are applied to suppress contributions from the decays $B^{0}\to K^{\ast 0} J/\psi (\to\!\ell^{+}\ell^{-})$, in which one of the leptons is misidentified and swapped with the $K^{+}$ or $\pi^{-}$. Signal yields are obtained with an unbinned maximum-likelihood fit to the distributions of the kinematic variable, $\Mbc = \sqrt{(E^{}_{\rm beam}/c^{2})^{2} - (p^{}_{B}/c)^{2}}$, where $E^{}_{\rm beam}$ is the beam energy and $p^{}_{B}$ is the momentum of the $B$ candidate in the center-of-mass frame.     

\begin{figure}[htb]
\includegraphics[width=0.325\textwidth]{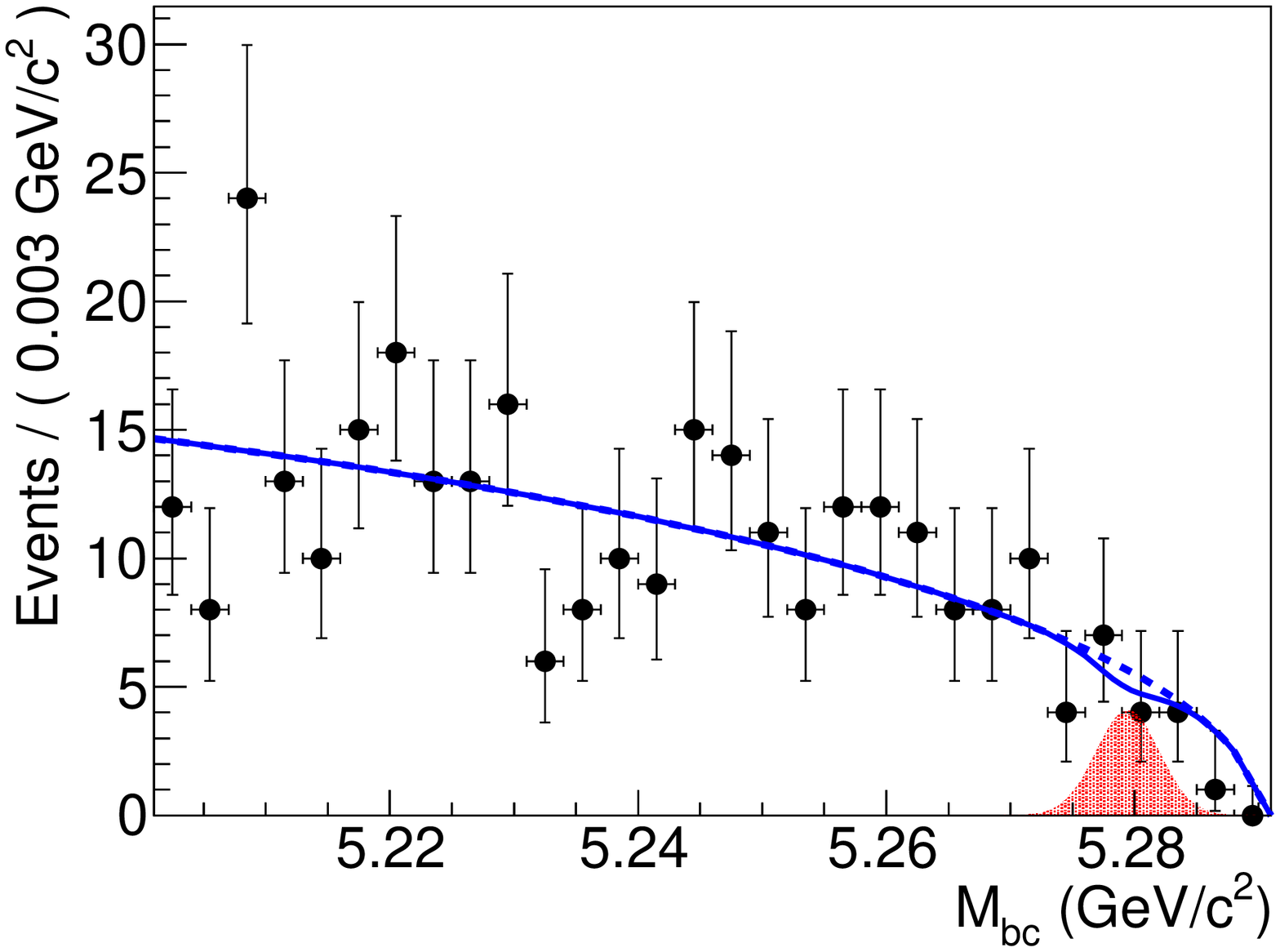}
\includegraphics[width=0.325\textwidth]{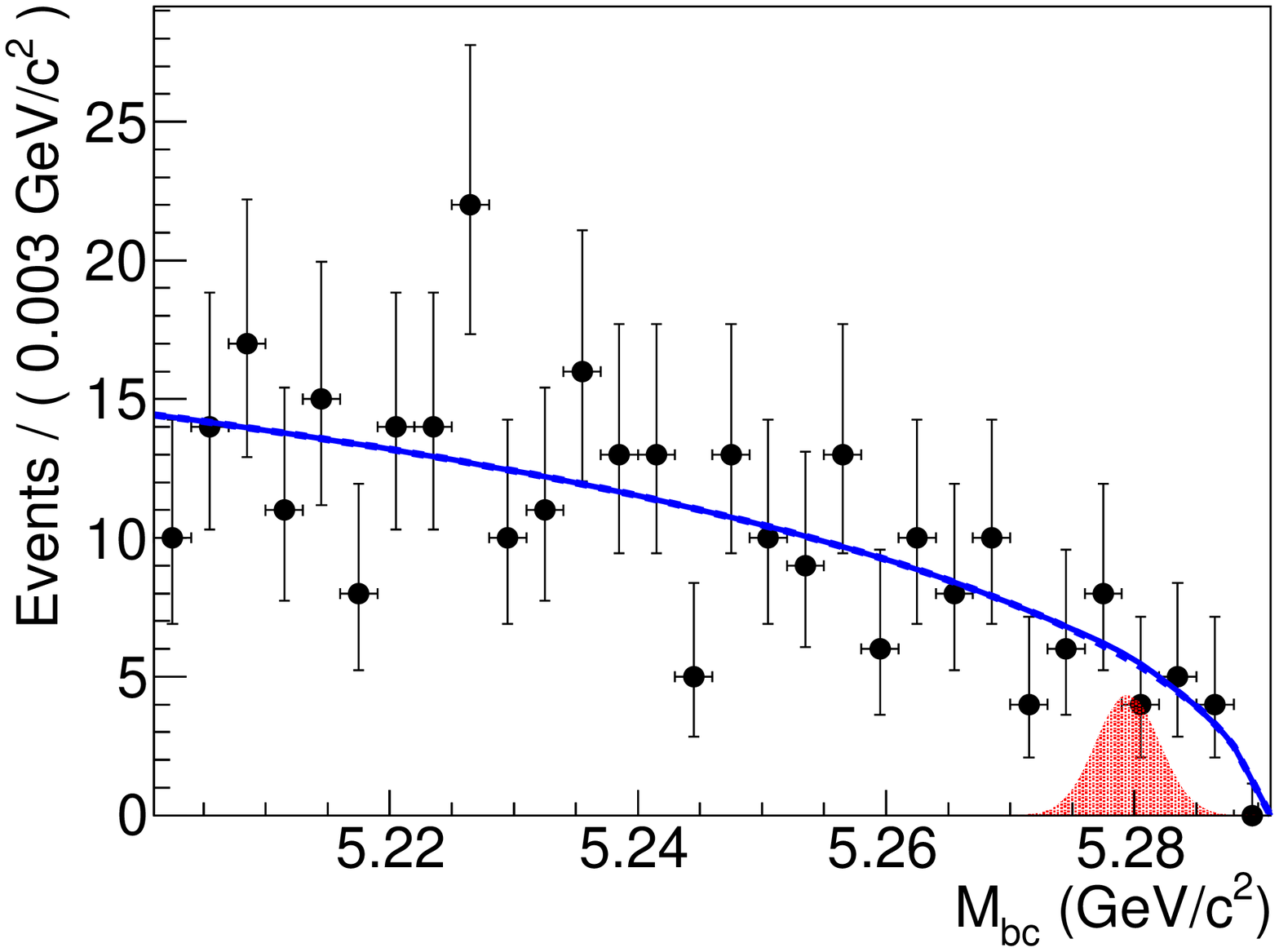}
\includegraphics[width=0.325\textwidth]{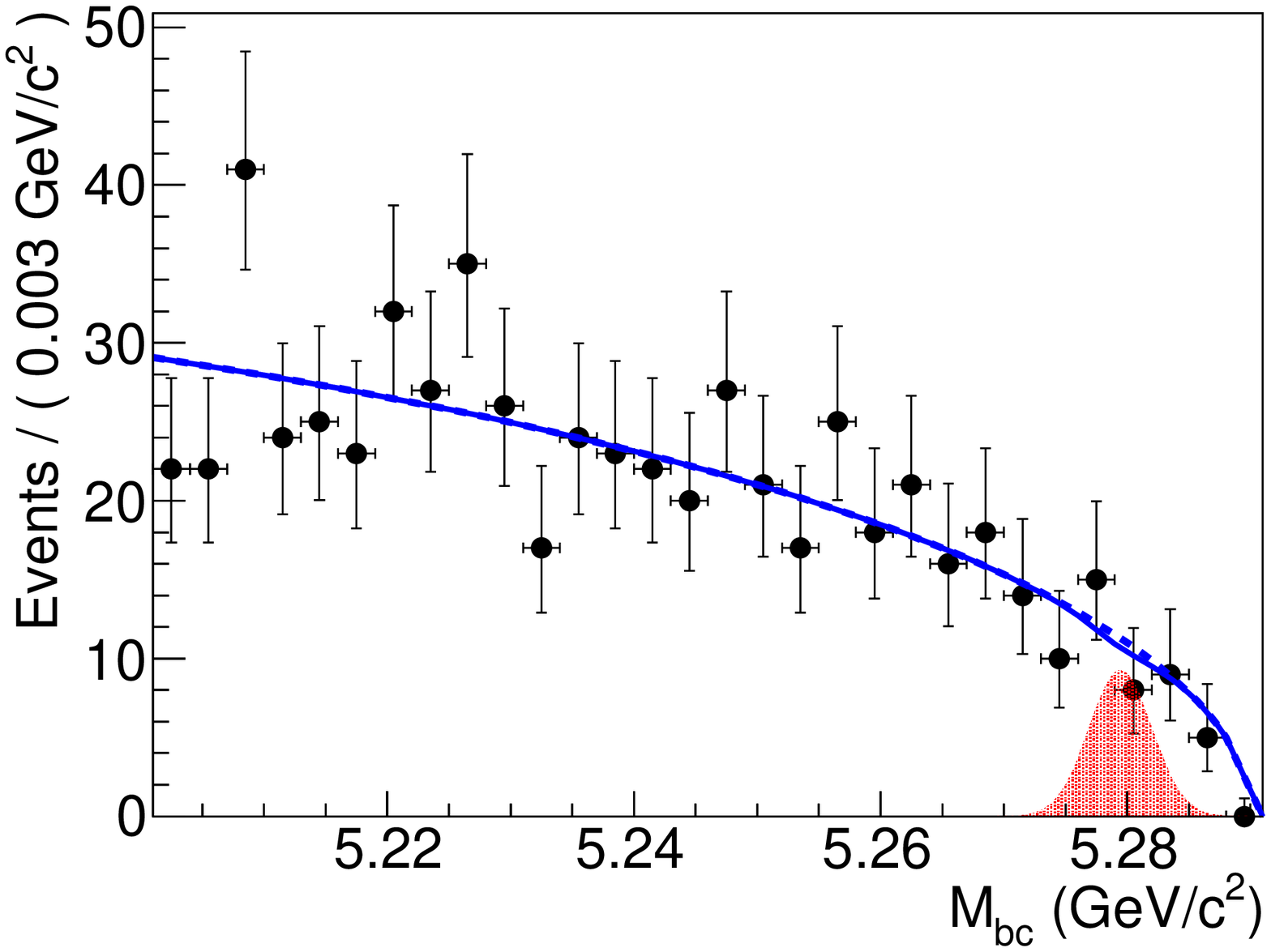}

\caption{The $\Mbc$ distributions for events that pass selection criteria for the decays $B^{0}\to K^{\ast 0} \mu^{+} e^{-}$ (left), $B^{0}\to K^{\ast 0} \mu^{-} e^{+}$ (middle), and also decays combined (right). The points with error bars are data, blue curve is the fit result, and the signal shape is depicted by the red shaded region with arbitrary normalization~\cite{36:ex:btokstmue}.}
\label{fig:btokstmue}
\end{figure}

As shown in Fig.~\ref{fig:btokstmue}, there is no evidence for signal due to the LFV decay. Therefore, upper limits at 90\% confidence level (CL) are set on $\pazocal{B}(B^{0} \!\to\! K^{\ast 0}\mu^{+}e^{-}) < 1.2 \times 10^{-7} $,  $\pazocal{B}(B^{0} \!\to\! K^{\ast 0}\mu^{-}e^{+}) < 1.6 \times 10^{-7}$, and on both the decays combined $\pazocal{B}(B^{0} \!\to\! K^{\ast 0}\mu^{\pm}e^{\mp}) < 1.8 \times 10^{-7}$. These are the most stringent limits on these decays to date. 

\section{Search for $B^{+} \to K^{+}\tau^{+}\tau^{-}$}
\label{btoktt}

The decay $B^{+} \to K^{+}\tau^{+}\tau^{-}$ is mediated via $b \to s\ell^{+}\ell^{-}$ FCNC process involving the third-generation lepton family, which can provide additional sensitivity to NP~\cite{37:th:btoktt}.   
First search for $B^{+}\to K^{+}\tau^{+}\tau^{-}$ is recently reported by the BaBar experiment~\cite{38:ex:btoktt}. In this study, the hadronic $B$-meson tagging method is used, where one of the two $B$ mesons produced in $\Y4S \to B^{+}B^{-}$ is reconstructed exclusively in many hadronic decay modes. The remaining tracks, clusters, and missing energy in the event is attributed to the signal $B$ meson. Only leptonic decays of the $\tau$ are considered, which results in three signal decay topologies with a $K^{+}$, multiple missing neutrinos, and either $e^{+}e^{-}$, $\mu^{+}\mu^{-}$, or $e^{+}\mu^{-}$. The neutrinos are account for the missing energy while any other neutral activity is discarded. Further, event shape variables are utilized to suppress continuum events. At this stage, the remaining backgrounds mostly arise from $B\bar{B}$ events, which are suppressed applying a criterion on the output of an NN formed with several input variables related to signal decay kinematics. No significant signal is observed and an upper limit on $\pazocal{B}(B^{+}\to K^{+}\tau^{+}\tau^{-}) < 2.25 \times 10^{-3}$ is obtained at the 90\% CL. 

\section{Search for $B \to h \nu \bar{\nu}$}
\label{btohnn}

The $B \to h \nu \bar{\nu}$ decays, (where $h$ refers to $K^{+}$, $K_{s}^{0}$, $K^{\ast +}$, $K^{\ast 0}$, $\pi^{+}$, $\pi^{0}$, $\rho^{+}$, or $\rho^{0}$) are FCNC processes with a neutrino pair in the final state. These FCNC decays involve the $Z$ boson alone, and hence are theoretically cleaner than $b \to s\ell^{+}\ell^{-}$ processes having a charged lepton pairs in the final state, where the photon also contributes~\cite{1:th:btohnn}.

\begin{figure}[htb]
  \centering
  \includegraphics[width=0.5\textwidth]{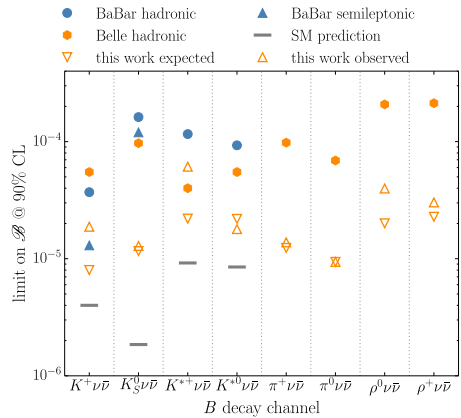}   
  \caption{Observed upper limits along with the expected values and previous measurement. SM predictions are also shown for the $K^{(\ast)}$ modes~\cite{btohnn:belle}.}
  \label{fig:btohnn}
\end{figure}

Previously, these decays were searched for in Belle utilizing the hadronic tagging method~\cite{btohnn:belle0} as well as in BaBar using both hadronic~\cite{btohnn:babar1} and semi-leptonic tags~\cite{btohnn:babar2}. The Belle analysis~\cite{btohnn:belle} is based on a more efficient semi-leptonic tagging method.
In this analysis, the signal $B$ daughter candidates are reconstructed through their subsequent decays: $K^{*0}\to K^{+}\pi^{-}$, $K^{\ast +}\to K^{+}\pi^{0}$ and $K_{s}^{0}\pi^{+}$, $\rho^{+}\to\pi^{+}\pi^{0}$, $\rho^{0}\to \pi^{+}\pi^{-}$, $K_{s}^{0}\to \pi^{+}\pi^{-}$, and $\pi^{0}\to \gamma\gamma$. Then, event shape variables are utilized to suppress the continuum background. Signal events are identified with the extra energy in the electromagnetic calorimeter, which is calculated by removing all the associated energy deposits due to tag and signal $B$ mesons. The largest signal contribution is observed in the $B\to K^{\ast +}\nu\nu$ decay with a significance of $2.3\sigma$.
In absence of a statistically significant signal in any of the decay modes, upper limits on their BFs are set at the 90\% CL. The result is summarized in Fig.~\ref{fig:btohnn} along with the expected values and previous measurements. These decays can be observed by Belle II with the uncertainties of similar size as that of current theoretical predictions~\cite{4:ex:belle2}.

\section{Summary}
\label{summ}
Decays involving $b\to s$ quark-level transitions are forbidden at tree level in the SM, but can proceed via penguin loop or box diagrams in which various NP particles may also contribute.
Belle reported the first evidence for isospin violation in the $B\to K^{\ast} \gamma$ decay~\cite{10:ex:btosg}; also first measurement of the difference of $CP$ asymmetries, between charged and neutral $B$ meson is performed in the same analysis.
In a sum-of-exclusive measurement of the decay $B \to X^{}_{s} \gamma$ at Belle, a null isospin asymmetry is found, which can reduce theoretical uncertainty in the BF; similarly, the $\Delta A^{}_{CP}$ value is found to be consistent with zero, helping constrain NP~\cite{9:ex:btosg}. 
The Belle's measurement of time-dependent $CP$ violation parameters in ${B^0 \to K_S^0 \eta \gamma}$~\cite{22:ex:btoksetag} is consistent with the null-asymmetry hypothesis within 2$\sigma$ as well as SM predictions.

An angular analysis for the decay $B\to K^{\ast}\ell\ell$ is performed~\cite{31:ex:btosll} for the first time in separate lepton-flavors; The results are consistent with both SM values and NP scenarios. As the measurement also hints at NP scenarios with possible LFU violation, this can eventually lead to LFV. The LFV decays $B^{0} \to K^{\ast 0} \mu ^{\pm} e^{\mp}$ are searched at Belle. The most stringent upper limit on the BF of these LFV decays are obtained with no evidence for any signal event~\cite{36:ex:btokstmue}.
The decay $B^{+}\to K^{+}\tau^{+}\tau^{-}$ is searched by BaBar for the first time and an upper limit on its BF is set at 90\% CL~\cite{38:ex:btoktt}.
Belle also reported a new search for the decay $B\to h\nu\nu$~\cite{btohnn:belle} based on a more efficient semi-leptonic tagging method. The obtained upper limits are close to SM predictions for the $K^{(\ast)}$ modes and Belle II has brighter prospects to observe these decays.


\begin{thebibliography}{99}
\bibitem {cc}
  The inclusion of charge conjugate decay modes is implied unless otherwise stated.
\end{thebibliography}
\end{document}